\documentclass{pasj00}
\draft

\begin{document}
\SetRunningHead{K. Asano, R. Yamazaki, and N. Sugiyama}
{UHECR from SGR 1806-20}
\Received{2005/July/29}
\Accepted{2005/December/9}

\title{Possibility of Ultra High-Energy Cosmic Rays from the Giant Flare in
SGR 1806$-$20}
\author{Katsuaki \textsc{Asano},\altaffilmark{1}
        Ryo \textsc{Yamazaki},\altaffilmark{2,3}
        and
        Naoshi \textsc{Sugiyama}\altaffilmark{1}
        }
\altaffiltext{1}{Division of Theoretical Astronomy,
              National Astronomical Observatory of Japan,\\
              2-21-1 Osawa, Mitaka, Tokyo 181-8588}
\altaffiltext{2}{Department of Earth and Space Science,
            Osaka University, Toyonaka, Osaka 560-0043}
\altaffiltext{3}{Department of Physics, Hiroshima University, 
            Higashi-Hiroshima, Hiroshima 739-8526}
\email{asano@th.nao.ac.jp, ryo@vega.ess.sci.osaka-u.ac.jp, naoshi@th.nao.ac.jp}


\maketitle

\begin{abstract}
On 2004 December 27, a giant flare from the soft gamma repeater
1806$-$20 was observed.
The radiation mechanism of the initial peak of the flare would be controversial.
In this letter we point out that very high-energy cosmic rays
would be produced in the case
that the flare was caused by internal shocks, as is usually
considered for gamma-ray bursts.
The highest energy of cosmic rays can reach $10^{19}$ eV,
if the Lorentz factor of the shocks is sufficiently high.
Future observations of cosmic rays will inform us
about the mechanism of the giant flare.
\end{abstract}

\section{Introduction}

Soft gamma repeaters (SGRs) are X-/gamma-ray transient
sources that show periods of bursting activity
separated by long intervals of quiescence.
They are galactic and LMC populations, and are
considered to originate from neutron stars
with intense ($\lesssim 10^{15}$ G) magnetic fields (magnetars)
\citep{td01}.

On 2004 December 27,
a giant flare from SGR 1806$-$20 was observed by several detectors
\citep{hur05,ter05,pal05,maz05}.
The initial peak of the giant flare had $\sim 0.6$ s duration.
The isotropic-equivalent energy would have been 
$E_\gamma\sim 10^{46{\mbox -}47} d_{15}{}^2$~erg
and the peak luminosity in the first 125 ms would have been 
$L_\gamma\sim 10^{47}d_{15}{}^2$~erg~${\mbox s^{-1}}$, 
where $d_{15}=d/15$ kpc, and $d$ is an uncertain distance to the source
\citep{cor04,cam05,mcc05}.

In the context of the magnetar model \citep{td01},
the giant flare 
arose from a hot expanding fireball,
which was only weakly polluted by baryons.
The initial optical depth for pair creation
is extremely large in this model.
As a result, this radiation-pair plasma expands relativistically.
The Lorentz factor, $\gamma$, increases in proportion to the radius, $R$, and
the temperature in the comoving frame decreases as $T \propto R^{-1}$
\citep{pir93}.
The photons decouple when the temperature decreases below $\sim 20$ keV 
\citep{nak05}.
At this stage the Lorentz factor becomes $\sim 10$, and
the radius becomes $R \sim 10 R_0$, where $R_0 \sim 10$ km is the size of the magnetar.
The photon temperature that we observe is the initial temperature independently of $\gamma$
at the decouple stage because of the relativistic blue-shift.
The predicted photon spectrum from this 
``pure radiation-pair fireball model'' is, therefore,
a quasi-blackbody with a temperature of $T\sim200$~keV \citep{hur05}.
The time-resolved (125 ms) energy spectrum from the {RHESSI}
particle detector is consistent with that of a black body
whose temperature is $\sim 200$ keV, though the main {RHESSI} spectroscopy
detectors were saturated during the peak \citep{bog05}.

However, there are so many ambiguities that one cannot confirm the
above picture.
The spectrum of the initial peak of the giant flare
is not yet well determined.
 \citet{hur05} reported on the cooling blackbody spectrum,
which is consistent with the pure radiation-pair fireball model.
However,
\citet{maz05} derived a  power-law spectral
shape ($\alpha=-0.7$) with an exponential cut-off at 800~keV.
There was another independent measurement, resulting in
a power-law spectrum ($\alpha=-0.2$) with an exponential
cut-off at 480~keV \citep{pal05}.
Furthermore, another giant flare in SGR~0526$-$66 may also have had a 
nonthermal spectrum \citep{fkl96}.
Therefore, we still have no definite evidence to interpret that
the flare on 2004 December 27 was a thermal one.

Another ambiguous point is the initial speed of the outflow 
or, almost equivalently, the  baryon richness.
The radio afterglow detected after the giant flare may give us
some hints to solve this problem \citep{cam05,gae05,tgg05}. 
\citet{wan05} showed that the same mechanism as established for
gamma-ray burst (GRB)
afterglows (decelerating-outflow model)
can explain the radio afterglow of the giant flare, suggesting
that the initial outflow was highly relativistic.
On the other hand, 
\citet{gra05} and \citet{gel05}
succeeded to fit the radio afterglow by their model,
in which the outflow is initially non-relativistic and
contains baryonic material of more mass than the $10^{24}$ g
ejected from the magnetar, itself.
The baryon amount in this picture, at first glance, may contradict
the optically thin emission of the flare.
Therefore, some authors \citep{eic05,dai05} propose multi-component models,
in which relativistic (baryon-poor) components
and non-relativistic (baryon-rich) components outflow from
the surface.

Jet collimation has been  suggested by
 \citet{yam05} (Y05) through reproducing
the observed light curve of the initial peak observed by {GEOTAIL}
with a 5.48~ms time resolution \citep{ter05}.
Using a simple emission model for relativistically moving matter
 \citep{yam03},
they derived an upper limit of the jet opening half-angle
of 0.2 rad.
In this model the initial peak in the giant flare
arises from internal shocks in relativistic jets.
The radius, $R_{\rm i}$, where the shock front starts to emit photons,
is estimated to be $R_{\rm i}=2.6 \times 10^8 \gamma^2$ cm,
which is much larger than the radius, $\gamma R_0 \sim 10^7$ cm,
derived from the pure radiation-pair fireball model.
Considering the opening half-angle of the jet, $\Delta \theta < 3/\gamma$
(Y05),
the collimation-corrected energy of the jet may be $10^{45}$ erg
if the radiation efficiency is assumed to be 10\%.
The observed proper motion of the centroid of a radio image \citep{tgg05}
may support the collimated outflow.
The relativistic jet is significantly decelerated due to 
the sideways expansion within ten minutes 
after the giant flare \citep{rho99,sar99,yam05}, so that
the radio image expands non-relativistically \citep{tgg05}.
Introducing a non-relativistic component besides the relativistic jet,
the observed expansion law of the radio image may also be
explained \citep{dai05}.

Therefore,
there are three controversial questions concerning this event:
1) Are there baryonic flows during the initial stage?
2) Are there relativistic components in the baryonic flows?
3) Does the flow have a jet-collimated structure?
We cannot, at present, definitely answer these questions.
In this paper we show that observations
of ultra-high-energy cosmic rays (UHECRs) or neutrinos
may resolve the above-mentioned problems \citep{eic03,iok05}.
\citet{iok05} extensively discussed the possibility of photopion production
and the resultant neutrino detection from this event.
On the other hand,
our discussion places emphasis on the detection possibility of
UHECRs delayed from the giant flare, based on the model in Y05,
which can explain the light curve of this event.

If there is a shocked outflow of baryons,
the shock waves may accelerate non-thermal protons to very high energies.
As for extragalactic GRBs, \citet{wax95,vie95} pointed out that
GRB is a candidate of UHECR sources, and
\citet{mil95} discussed possible
observations of UHECRs.
In this paper we discuss the possibility of UHECR production in the giant flare
of SGR 1806$-$20.
Combining neutrino observations, UHECR detections
may provide information on the above questions 1) and 2).
In section 2 we obtain the maximum energy of UHECRs accelerated
in shock waves based on the jet model of Y05.
In section 3 the expected number flux of UHECRs, which depends on
the structure of the galactic magnetic fields, is discussed.
Our conclusions are summarized in section 4.

\section{Maximum Energy of Cosmic Rays}

In this section we adopt the jet model of Y05 and
estimate the maximum energy of cosmic rays that we can observe.
We expect that even another internal shock model may predict similar
maximum energies to ours, as long as the assumed energy and time-scale of the flare
are close to ours (see e.g. Eichler 2005).

In the model of Y05 a shell moving with the Lorentz factor, $\gamma$,
emits gamma rays from
$R=R_{\rm i}=2.6 \times 10^8 \gamma^2$ cm to $R_{\rm e} \simeq 10 R_{\rm i}$.
The duration in the comoving frame is
$(R_{\rm e}-R_{\rm i})/(c \gamma) \sim R_{\rm e}/(c \gamma)$,
which corresponds to the shell crossing time scale.
Therefore, the shell width in the observer frame is about $R_{\rm e}/\gamma^2$,
while the shell width used conventionally in the GRB models
is $R_{\rm i}/\gamma^2$, which implies $R_{\rm e} \sim R_{\rm i}$.
Thus, the shell width obtained from the {GEOTAIL} light curve
is relatively thick.
As a result, the physical conditions around $R \sim R_{\rm i}$ and
$R \sim R_{\rm e}$ are different.

It is difficult to estimate the magnetic field in the shell,
because the light curve provides information only on the kinematic properties.
The plasma in the shell coming from the magnetar may be strongly magnetized.
A part of the energy of the magnetic field may be expended
in the acceleration phase, while various plasma instabilities
may enhance the field, as is discussed by \citet{med99} for GRBs.
Though we recognize the above difficulties,
we assume that the energy density of the magnetic field
is proportional to the photon energy density.
The luminosity, which is Lorentz invariant,
is $E_\gamma c/R_{\rm e}=1.2 \times 10^{43} E_{46} \gamma_2^{-2}$
erg ${\mbox s^{-1}}$,
where $E_\gamma=10^{46} E_{46}$ erg.
In this case the photon energy density in the comoving frame
is approximated as $U_\gamma=E_\gamma/(4 \pi R^2 R_{\rm e})$.
From $B^2/(8 \pi)=\varepsilon_{\rm B} U_\gamma$,
we obtain $B(R)=1.1 \times 10^{4} \varepsilon_{\rm B}^{1/2}
E_{46}^{1/2} \gamma_2^{-3} (R/R_{\rm i})^{-1}$ G,
where $\gamma=100 \gamma_2$.
The energy ratio, $\varepsilon_{\rm B}$, can be larger than unity,
if the baryonic energy density is much larger than $U_\gamma$,
which implies that the gamma-ray emission is an inefficient process.

The shocks may accelerate non-thermal protons.
The maximum energy of accelerated particles, $E_{\rm max}$,
is restricted by the condition that the particle Larmor radius,
$r_{\rm L}=E/\gamma e B$,
is smaller than the size scale of the emitting region,
$(R-R_{\rm i})/\gamma \sim R/\gamma$.
This condition gives the maximum energy,
$E_{\rm L,max}=8.3 \times 10^{18} \varepsilon_{\rm B}^{1/2}
\gamma_2^{-1} E_{46}^{1/2}$ eV, which is independent of $R$.

Another condition to generate cosmic rays is that
the cooling time scale of protons should be longer than
the dynamical time scale $T_{\rm dyn}=R/(c \gamma)$.
As long as we consider energies below $E_{\rm L,max}$,
the acceleration time scale ($\sim r_{\rm L}/c$) may be shorter
than $T_{\rm dyn}$.
If the protons accelerated to high energies cool down
before they escape from the shell,
those particles cannot become cosmic rays.
For proton-synchrotron cooling we obtain the maximum energy
of $E_{\rm syn,max}=4.6 \times 10^{21} \varepsilon_{\rm B}^{-1}
\gamma_{2}^6 E_{46}^{-1} (R/R_{\rm i})$ eV.

We should comment on the ``adiabatic cooling'' due to
shell expansion.
While the shell coasts from $R_{\rm i}$ to $R_{\rm e}$,
the shocked region grows via shock propagation.
The adiabatic invariance, $B r_{\rm L}^2$, may lead
to the cooling of particles.
However, we are not sure whether the adiabatic condition
is satisfied or not for a disturbed magnetic field in the
shocked region.
In any case, we assume that particles escape from the shell
on the dynamical timescale at each radius via stochastic processes.
Therefore, we neglect adiabatic cooling in our case.

The maximum energy of cosmic rays is determined by
min$( E_{\rm L,max},E_{\rm syn,max} )$.
The results are plotted in figure 1.
For a smaller $R$ the maximum energy
is determined by $E_{\rm syn,max}$,
because the magnetic field becomes stronger.
As $R$ increases the maximum energy is determined by
$E_{\rm L,max}$, rather than $E_{\rm syn,max}$,
because the weaker magnetic field cannot confine high-energy
particles in the shell.
For $\gamma \lesssim 30$ the radius where the shock occurs ($\propto \gamma^2$)
is smaller, so that the energy limit is determined
by $E_{\rm syn,max}$ only between $R_{\rm i}$ and $R_{\rm e}$.
On the other hand, for $\gamma \gtrsim 40$, the limit is determined
by only $E_{\rm L,max}$.
The values, $E_{\rm max}$, achieved in shocked shells with $\gamma$
are the maximum values in each line in figure 1.
For $\varepsilon_{\rm B}=1$ and $E_{46}=1$,
$E_{\rm max}$ may be a few times $10^{19}$~eV.

\begin{figure}
   \includegraphics[width=7cm]{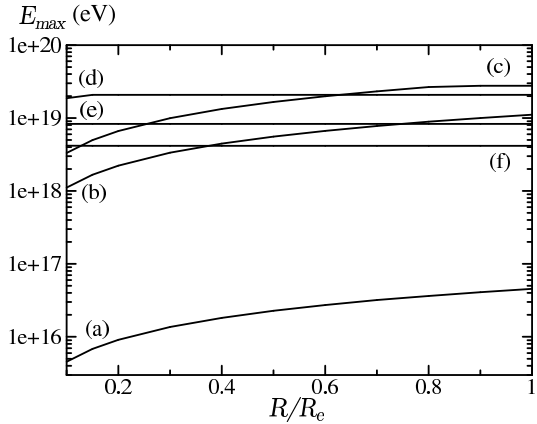}
\caption{Maximum energy, $E_{\rm max}$, determined by
$E_{\rm L,max}$ and $E_{\rm syn,max}$
for $\varepsilon_{\rm B}=1$ and $E_{46}=1$.
The Lorentz factor, $\gamma$, is assumed to be (a) 10,
(b) 25, (c) 30, (d) 40, (e) 100, and (f) 200.}
\end{figure}

Another cooling process that we should take into account
is photopion creation.
If the proton photopion ``optical depth'' is much higher
than unity, protons lose their energies in the flare source
before they escape from the emitting region \citep{asa03,asa05}.
The condition to create pions is 
 $E_{\rm CR} \epsilon_\gamma \geq 0.2~ {\rm GeV}^2$, where
 $E_{\rm CR}$ and $\epsilon_\gamma$ are the
energies of a nucleon and an interacting photon, respectively.
In order to estimate the energy loss rate due to photopion creation,
we need the soft photon spectrum in the flare, which is unknown.

If the giant flare is thermal emission of $\sim 200$ keV,
the number of soft photons
that interact with protons of $\sim 10^{19}$ eV is too small to
create pions on the dynamical time scale.
However, it may be premature to conclude that the spectrum observed by
{RHESSI} is thermal.
As suggested in \citet{maz05} and \citet{pal05},
there is a possibility that soft photons due to non-thermal electrons
may distribute below $\sim 200$ keV with the spectrum
$n(\epsilon_\gamma) \propto \epsilon_\gamma^{-p}$,
as observed in standard GRBs.

As the most pessimistic case,
we assume that power-law photons with $p=1$ or $1.5$ dominate
below 200 keV without the low-energy cut-off.
Above $200$ keV we adopt the Planck spectrum with $200$ keV,
though the spectrum shape above 200 keV is not important
for the cooling process of ultra-high-energy particles.
Although the number of power-law photons is much
larger than the high-energy photons,
the energy contribution of the soft photons is not important for $p<2$.
Using the same cross section and method as \citet{asa05},
we estimate the time scale of the photopion creation, $T_{\rm pi}$
near $R_{\rm max}$.
The results are plotted in figure 2.

\begin{figure}
   \includegraphics[width=7cm]{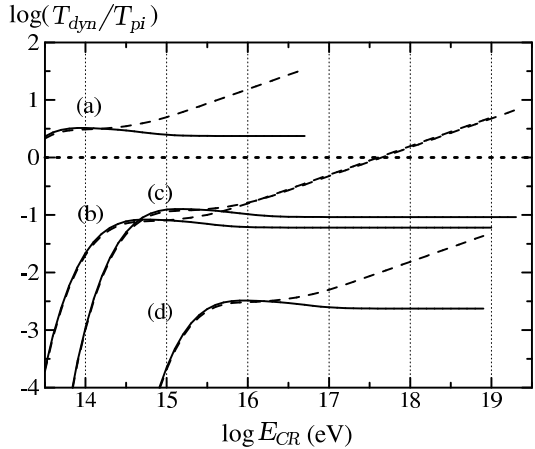}
\caption{Time scale ratio $T_{\rm dyn}/T_{\rm pi}$
vs. energy of nucleons for photon index $p=1$ ({solid})
and $p=1.5$ ({dashed}). The model parameters $(\gamma,R/R_{\rm i})$
are (a) $(10,10)$, (b) $(25,10)$, (c) $(40,1)$, and (d) $(100,1)$.
The total photon energy, $E_\gamma$, is fixed as $10^{46}$ erg.
The dashed lines of (b) and (c) nearly overlap.
}
\end{figure}

The low-energy bump appearing in figure 2. is due to the ``thermal'' photons
above $200$ keV.
Above this energy nucleons interact with
the power-law photons.
In this case the number of interacting photons
($n(\epsilon_\gamma) d \epsilon_\gamma$, $\epsilon_\gamma \sim
0.2 \gamma^2 {\rm GeV}^2/E_{\rm CR}$)
is independent of $E_{\rm CR}$ for $p=1$.
As a result, the time scale is nearly constant in this energy region.
For $p=1.5$ the time scale becomes shorter as $E_{\rm CR}$ increases
($T_{\rm pi} \propto E_{\rm CR}^{-1/2}$).
If the power-law spectrum has
a low-energy cut-off due to synchrotron self-absorption etc.,
the time scale will increase from the corresponding energy.

For $\gamma=10$ the photopion production is crucial,
and resultant high-energy neutrinos will be emitted as is the case in \citet{iok05}.
However, if $\gamma$ is large enough to generate UHECRs,
the photopion production is inefficient.
Even for $\gamma=25$, UHECRs will survive and neutrinos may not be observed,
unless the photon spectra are extremely soft, such as $p=1.5$.

Unless the energy of soft photons below $50$ keV is much larger
than the observed energy by {RHESSI},
there is a possibility that UHECRs of $>10^{19}$ eV would come from
the SGR 1806$-$20 giant flare.
At  present, considering the ambiguity in the parameter values,
we cannot completely preclude the detection possibility of UHECRs
above $10^{20}$ eV
(for example, $\varepsilon_{\rm B}=10$, $\gamma=30$, and $E_{46}=3$).

\section{Propagation of UHECRs}

The source location of the flare is about $10^\circ$ from the galactic center.
In the region around the galactic center the magnetic field is highly uncertain.
If the magnetic fields are well represented by regular fields along the
spiral arms, the time delay may be on the order of $\sim 10$~yrs,
even for $10^{20}$ eV \citep{alv02}.
Turbulent magnetic fields, whose scale is 10--100 pc \citep{bec01},
may shorten the above estimate.
In the most pessimistic case, the time delay becomes on the order
of thousands of yrs \citep{eic05}.

If the energy of cosmic rays above $10^{19}$ eV is $P_{19}$ \%
of the total photon energy $E_\gamma \sim 10^{46}$ erg,
$2 P_{19} d_{15}^{-2} \Delta T_3^{-1}$ particles are detected per one year by
1000 ${\rm km}^{2}$ detectors, such as {AUGER} or Telescope Array,
where $\Delta T_3$ is the dispersion of the timedelay
normalized by 1000 yr.
If the particles are confined to 0.1 rad of the galactic plane,
the above detection rate can be sufficient signals, even for $\Delta T_3=1$
\citep{eic05}.

\section{Conclusions and Discussion}

If the giant flare from SGR 1806$-$20 on 2004 December 27
is due to internal shocks in relativistic jets,
we found that the shocks may produce UHECRs of up to $\sim 10^{19}$ eV.
The maximum energy is similar to an estimate by \citet{eic05}.
In standard GRBs, on the other hand,
UHECRs are unlikely to be observed, since they lose energy before escaping
from the shell via photopion production \citep{asa05}.
Even if UHECRs are produced in GRBs,
the time delay between cosmological bursts and UHECRs
is too large to establish any connection between the GRBs and UHECRs.
On the other hand,
because the giant flare of SGRs is less luminous than standard GRBs,
UHECRs can escape from the shell without losing their energy.

In our baryon-loaded jet model, if the bulk Lorentz factor of the outflow is
high enough, neutrinos that are produced
from the decay of charged pions may not come from this flare.
This is consistent with the results of \citet{iok05}.
On the other hand, if the outflow is non-relativistic,
neutrinos could be produced via p-p collisions,
which is neglected in our calculation,
in addition to photopion production.
Hence, UHECRs and high-energy neutrino observations become
diagnostic tools to investigate the properties of the outflows
at an early phase.
There are three possible cases:
A) detection of neutrinos,
B) detection of UHECRs, but no neutrinos, and
C) no detection of high-energy particles.
We may reach the following conclusion irrelevant of the models.
Cases A and B mean there are baryons in the flare stage, but
the Lorentz factor is not very large in case A (see also Gelfand et al. 2005).
Case C implies that there are negligible baryons 
(pure radiation-pair fireball model),
or the UHECR production efficiency is low.

Because the arrival time of UHECRs depends on a number of factors,
it cannot be predicted exactly.
It might be at present that 
we will detect UHECRs coming from the other two SGRs
(SGR~0526$-$66, SGR~1900$-$14) that previously caused giant flares.
Giant flares of SGRs could produce a large amount of UHECRs,
which may explain the origin of doublet and/or triplet events,
though there is no evidence of any correlation between such events
observed by {AGASA} and SGRs so far.
Anomalous X-ray pulsars (AXPs) may be the same kind of objects as SGRs and
older, but less active than SGRs \citep{kul03}.
They might have produced UHECRs via giant flares thousands of years ago.
We could detect signals from past activities of AXPs by UHECR observations.

\vspace{1.5cm}

This work was supported in part by 
JSPS Research Fellowship for Young Scientists (RY).

\end{document}